\documentstyle[12pt,geompsfi,epsf]{article}
\newcommand{\bea}{\begin{eqnarray}}
\newcommand{\eea}{\end{eqnarray}}
\newcommand{\bean}{\begin{eqnarray*}}
\newcommand{\eean}{\end{eqnarray*}}
\newcommand{\ba}{\begin{array}}
\newcommand{\ea}{\end{array}}
\newcommand{\be}{\begin{equation}}
\newcommand{\ee}{\end{equation}}
\newcommand{\nn}{\nonumber}

\newcommand{\bra}[1]{\langle #1|}
\newcommand{\ket}[1]{|#1\rangle}
\newcommand{\av}[1] {\langle #1\rangle}

\newcommand{\amp}[2]{\langle #1|#2\rangle}

\newcommand{\calD}{\mbox{${\cal D}$}}

\newcommand{\calO}{\mbox{${\cal O}$}}

\newcommand{\Tr}{\mbox{${\rm Tr~}$}}
\newcommand{\tr}{\mbox{${\rm tr~}$}}
\newcommand{\pa}{\partial}
\newcommand{\da}{\dagger}
\newcommand{\ra}{\rightarrow}
\newcommand{\lr}{\leftrightarrow}
\newcommand{\phistar}{\phi^*}
\newcommand{\bp}{{\bf p}}
\newcommand{\bk}{{\bf k}}
\newcommand{\bx}{{\bf x}}
\newcommand{\by}{{\bf y}}
\newcommand{\bz}{{\bf z}}
\newcommand{\bA}{{\bf A}}
\newcommand{\bE}{{\bf E}}
\newcommand{\hp}{\hat \psi}
\newcommand{\hpd}{\hat \psi ^\dagger}
\newcommand{\bfga}{\mbox{\boldmath $\gamma$}}
\newcommand{\bfsi}{\mbox{\boldmath $\sigma$}}
\newcommand{\tf}{t_f}
\newcommand{\hatN}{\hat{N}}

\newcommand{\fd}[1]{\frac{\delta }{\delta #1}} 
\newcommand{\pad}[1]{\frac{\partial}{\partial #1}} 
\newcommand{\refpa}[1]{(\ref{#1})} 

\newcommand{\barm}{\bar{m}}
\newcommand{\barom}{\bar{\omega}}
\newcommand{\barbeta}{\bar{\beta}}
\newcommand{\barp}{\bar{p}}
\newcommand{\barbp}{\bar{{\bf p}}}
\newcommand{\bart}{\bar{t}}
\newcommand{\bartf}{\bar{t}_f}
\newcommand{\barV}{\bar{V}}
\newcommand{\barbe}{\bar \beta}
\newcommand{\om}{\omega}
\newcommand{\Om}{\Omega}

\newcommand{\prd}[3] {Phys.\ Rev.\ D               {#1} {(#2)} {#3}}
\newcommand{\annp}[3]{Ann.\ Phys.\ (N.Y.)          {#1} {(#2)} {#3}}
\newcommand{\pr}[3]  {Phys.\ Rev.                  {#1} {(#2)} {#3}}
\newcommand{\npb}[3] {Nucl.\ Phys.\ B              {#1} {(#2)} {#3}}
\newcommand{\prl}[3] {Phys.\ Rev.\ Lett.           {#1} {(#2)} {#3}}

\begin{document}
\begin{titlepage}
\begin{flushleft}
G\"oteborg\\
ITP 94-40\\
hep-th/9412188\\
December 1994\\
\end{flushleft}
\vspace{0.5cm}
\begin{center}
\vspace{1.5cm}

{\large \bf
 Fermionic and bosonic pair creation in an external electric
field at finite temperature
using the functional Schr\"odinger representation}\\
\vspace{5mm}
{  Joakim Hallin\footnote{Email address: tfejh@fy.chalmers.se}
and Per Liljenberg\footnote{Email address: tfepl@fy.chalmers.se}}\\
\vspace{1cm}
{\sl Institute of Theoretical Physics\\
Chalmers University of Technology\\
and University of G\"oteborg\\
S-412 96 G\"oteborg, Sweden}\\

\end{center}
\begin{abstract}
We solve the time evolution of the density matrix both for fermions and bosons
in the
presence of a homogeneous but time dependent external electric field. The
number of particles produced by the external field, as well as their
distribution in momentum space is
found for finite times. Furthermore, we calculate the probability of finding a
given
number of particles in the ensemble. In all cases, there is a
nonvanishing thermal contribution. The bosonic
and the fermionic density matrices are expressed in a "functional field basis".
This constitutes an extension of the "field basis" concept to fermions.
\end{abstract}
\end{titlepage}

\section{Introduction}

One of the characteristic features of quantum field theory is the
non-conservation of the number of particles.
 The most familiar cases are the collision processes,
where genuinely quantum mechanical interactions
lead to creation (or annihilation) of particles. Here, all the interacting
fields are
quantum fields. It has been known for long that also classical external fields
can
change the number of particles. This is the well-known Schwinger mechanism for
particle production in an external field \cite{schwinger}.
The Schwinger formula
has been derived in a number of ways and has found a wide range of
applications, from Hawking radiation \cite{brout} to heavy ion collisions
\cite{sailer}.

Schwinger's result has some shortcomings. The external field is
homogeneous throughout space; the timescale is large (infinite) and it applies
only at
zero temperature. Furthermore the probability of finding the system in
the groundstate after
some time has elapsed is
not
related in a simple manner
to observable quantities, like the average number of particles
created because the probability distribution is non-trivial.

The first problem has to some extent been discussed elsewhere \cite{wang} and
will not
be considered here.

The problem of large times comes about
 because of the nature of the model, which consists of
free particles minimally coupled to
an external background field. In this case the backreaction, i.e. the field
generated by created particles is neglected. Since the number of
 particles, and hence the induced field, increases with time there is a maximal
time for which the background field approximation applies. If one insists
on that approximation, which we do,
 it is important to keep the explicit time
dependence throughout the calculations.
 The backreaction has been studied extensively in
\cite{mottola}.

The third problem of extending the
pair creation phenomenon to non-zero temperatures has been studied in thermal
field theory, where the effective action is a natural object of interest.
However, there seems to
be confusion about the influence of a thermal background on the pair
creation \cite{elmfors,loewe,cox}. In the real time formalism one does
not find a thermal correction
whereas using imaginary time there appears to be one. It is also not clear
if any of the formalisms actually are applicable to this kind of problems.
In this context it is again  necessary to keep
 the time dependence and care must be taken with the infinite time limit.
We find that at fixed temperature,
 the thermal contribution to pair production will vanish in the infinite time
limit.
At any finite time, however, there is a significant thermal effect. In the
background
field approximation, one can therefore question the use of effective
lagrangians
(the infinite time limit being understood)
at least concerning the imaginary part responsible for pair production.

So much for the physics, we now turn to the formalism. We will be working in
the functional Schr\"odinger representation (see e.g.
\cite{jackiw,kiefer92,kiefer94})
which is convenient for many reasons. Because of the close
analogy with quantum mechanics, there are no conceptual
problems in this picture. We do not suffer from doubts about the
formalism and/or applicability, as in thermal field theory. Furthermore,
the treatment of time and temperature is straightforward.

In order to make the calculations explicit we make extensive use of the
so called "field basis" concept, i.e. one views the wave functionals as
overlaps of a general state and a "field basis state". This is known
to be possible for bosons. We extend the formalism to fermions.
One then has to be careful with the Grassmann nature of the basis. Also extra
care must be taken when working in a reducible representation, which is the
case at hand.

With this setup, the strategy is simple.
We solve the time evolution for the thermal density matrix of the
system under consideration,
i.e. we find $\rho$ satisfying
\be
i\dot{\rho}=[H,\rho]~~,~~\rho(0)={\rm equilibrium\  ensemble}.
\ee
Then we simply calculate the thermal average of the number operator $N$, that
is
\be
\av{N} = \tr[N \rho(t)] .
\ee
Closely related to this quantity is the number density in momentum space
(transverse
and longitudinal momenta). We also find the probability distribution of the
n-particle states in the
ensemble at arbitrary times, specifically we calculate the probability of
finding
the system in the groundstate at finite times.
The paper consists of two main sections. In the first section we treat
bosons and set up the machinery.
Fermions are dealt with in the second section. After a careful discussion of
the
fermionic field basis, all the calculations previously done for bosons
are repeated for fermions. The paper ends with a few concluding remarks.

A word of caution; throughout this paper there are numerous seemingly infinite
quantities. This is not case however, since we always work with a UV-cutoff.
Only
in the expressions that are not only convergent but also quite insensitive to
the
cutoff, do we remove this cutoff.

\section{Bosons}
The hamiltonian for a charged scalar field in an external electromagnetic field
is
\be\label{e:hamiltonian}
\hat{H} = \int  d^3x \hat{\pi}(\bx) \hat{\pi}^\dagger(\bx) + \int  d^3x
d^3y\hat{\phi}^\dagger(\bx) \omega^2(\bx,\by) \hat{\phi}(\by),
\ee
where
\be
\omega^2(\bx,\by)= (-(\nabla_x-ie\bA(\bx))^2+m^2) \delta(\bx-\by)
\ee
and we use the convention that $e=-|e|<0$. In a standard manner
\cite{jackiw,kiefer92}
we represent the field operators and the states:
\bea
 \hat{\phi}\lr\phi~,~\hat{\pi}\lr\frac{1}{i}\frac{\delta}{\delta \phi}~&,&~
\hat{\phi}^\dagger\lr\phistar~,~\hat{\pi}^\dagger\lr\frac{1}{i}\frac{\delta}
{\delta \phistar}\\
 \ket{\Psi}\lr\Psi(\phi,\phistar)~&,&~ \bra{\Psi}\lr\Psi(\phi,\phistar)^*.
\eea
By $*$ we mean complex conjugation.
 The inner product is defined by functional integration
\be
\amp{\Psi}{\Phi}=\int \calD^2\phi  \Psi(\phi,\phistar)^*\Phi(\phi,\phistar),
\ee
where $\calD^2\phi=\calD\phi\calD\phi^*$. As a reference we give a formula for
gaussian functional integrals:
\bea
\lefteqn{\int  \calD^2\phi \exp{\left[-\int d^3x d^3y ~\phistar\Omega\phi+\int
d^3x ~(a^*\phi+\phistar b)\right]}=}\nn  \\
& &~~~~~~~~~~~~~~~~~~~~~~~~~~~~~=\det(\Omega)^{-1} \exp{\left[\int d^3x d^3y
{}~a^*\Omega^{-1}b\right]},
\eea
where the determinant is defined through $\det \Omega = \exp \Tr \ln \Omega$
with a functional logarithm and $\Tr \Omega =\int d^3x ~\Omega(\bx,\bx)$.

The state functional can be viewed as an overlap between a general
state and a "field basis state", thus
\be
\Psi(\phi,\phistar)=\amp{\phi,\phistar}{\Psi}.
\ee
The field basis has the following properties \cite{jackiw}:
\bea
\hat{\phi}\ket{\phi,\phistar}&=&\phi\ket{\phi,\phistar},\\
\amp{\phi,\phistar}{\varphi,\varphi^*}&=
&\delta(\phi-\varphi)\delta(\phistar-\varphi^*),\\
\hat{1}&=&\int \calD^2\phi\ket{\phi,\phistar}\bra{\phi,\phistar}.
\eea
For operators we have
\bea
\bra{\phi,\phistar}\hat{\calO}\ket{\varphi,\varphi^*}&=&
\calO(\phi\phistar,\varphi\varphi^*),\\
\tr \hat{\calO} &=& \int \calD^2\phi
\bra{\phi,\phistar}\hat{\calO}\ket{\phi,\phistar},\\
\bra{\phi,\phistar}\hat{\calO}(\hat{\phi},\hat{\phi}^\dagger,\hat{\pi},
\hat{\pi}^\dagger)
\ket{\Psi}&=&\hat{\calO}(\phi,\phistar,\frac{\delta}{\delta \phi},
\frac{\delta}{\delta \phistar})\amp{\phi,\phistar}{\Psi}.
\eea

\subsection{The initial ensemble}

The system under consideration is described by a density matrix $\rho(t)$. For
a good
introduction to density matrices see \cite{feynman}. We intend to
evolve the system in time, using the hamiltonian \refpa{e:hamiltonian}.
The system evolves
from an initial ensemble at thermal equilibrium at $t=0$ to a final ensemble
at $t=\tf$ when the external field
is switched off and the number of particles in the system is counted.

Initially the system is in thermal equilibrium and is described by an
unnormalized
stationary density matrix $\hat{\rho}_u=e^{-\beta \hat{H}_0}$ satisfying
\be
-\pa_\beta\hat{\rho}_u=\hat{H}_0\hat{\rho}_u \label{e:ini1}
\ee
where $\hat{H}_0$ is the hamiltonian \refpa{e:hamiltonian}
with $\bA=0$ and $\beta$ is the inverse temperature, $\beta=\frac{1}{k_B T}$.
Note that
such an ensemble cannot be created by pure QED since it contains all possible
charges.
Furthermore we have
\be
\lim_{\beta \ra
0}\hat{\rho}_u=\hat{1}~~,~~\hat{\rho}_u^\dagger=\hat{\rho}_u.\label{e:ini2}
\ee
In the functional representation $\hat{\rho}_u$ is represented by a functional
matrix
$\rho_u(\phi_1\phistar_1,\phi_2\phi_2^*)=
\bra{\phi_1,\phistar_1}\hat{\rho}_u\ket{\phi_2,\phi_2^*}$, thus
\be\label{e:init}
-\pa_\beta\rho_u(\phi_1\phistar_1,\phi_2\phi_2^*)=
\hat{H}_0(\phi_1,\phistar_1,\frac{\delta}{\delta \phi_1},\frac{\delta}{\delta
\phi_1^*})
\rho_u(\phi_1\phistar_1,\phi_2\phi_2^*),
\ee
and the two additional conditions read:
\bea\label{e:delta}
\lim_{\beta \ra
0}\rho_u(\phi_1\phistar_1,\phi_2\phi_2^*)&=&\delta(\phi_1-\phi_2)
\delta(\phi_1^*-\phi_2^*)\\
\rho_u(\phi_1\phistar_1,\phi_2\phi_2^*)&=&\rho_u^*(\phi_2\phi_2^*,
\phi_1\phistar_1) \label{e:herm}.
\eea
In order to solve for $\rho_u$, we make a gaussian ansatz with a covariance
$\Omega^0_{ij}$
\be
\rho_u(\phi_1\phistar_1,\phi_2\phi_2^*)=N_u \exp{\left[-\int  d^3x
d^3y\phistar_i(\bx)\Omega^0_{ij}(\bx,\by)\phi_j(\by)\right]}~~~i,j=1,2.
\ee
One obtains
\bea
-\pa_\beta \ln N_u &=& \Tr \Omega^0_{11} \\
-\pa_\beta \Omega^0_{11}&=&{\Omega^0_{11}}^2-\omega_0^2\\
-\pa_\beta \Omega^0_{12}&=&\Omega^0_{11}\Omega^0_{12} \\
-\pa_\beta \Omega^0_{21}&=&\Omega^0_{21}\Omega^0_{11} \\
-\pa_\beta \Omega^0_{22}&=&\Omega^0_{21}\Omega^0_{12}
\eea
where $\om_0=\om$ having set $\bA=0$. We have also used the notation
\be
(\Omega \tilde{\Omega})(\bx,\by)=\int d^3z
\Omega(\bx,\bz)\tilde{\Omega}(\bz,\by).
\ee
Due to translational invariance the covariances will satisfy
$\Omega^0_{ij}(\bx,\by)=\Omega^0_{ij}(\bx-\by)$. We can therefore simplify the
equations by performing a fourier transform according to
\be
\phi(\bx)=\int_{p} \phi(\bp) e^{i \bp\cdot \bx}~,~\pi(\bx)=\int_{p} \pi(\bp)
e^{-i \bp\cdot \bx}~,~\Omega(\bx-\by)=\int_{p} \Omega(\bp) e^{i \bp\cdot
(\bx-\by)}
\ee
where $\int_{p}=\int d^3\bp/(2\pi)^3$. The equations then read
\bea\label{e:initial1}
-\pa_\beta \ln N_u &=& V \int_p \Omega^0_{11}(\bp) \\
-\pa_\beta \Omega^0_{11}(\bp)&=&{\Omega^0_{11}(\bp)}^2-\omega_0(\bp)^2\\
-\pa_\beta \Omega^0_{12}(\bp)&=&\Omega^0_{11}(\bp)\Omega^0_{12}(\bp) \\
-\pa_\beta \Omega^0_{21}(\bp)&=&\Omega^0_{21}(\bp)\Omega^0_{11}(\bp) \\
-\pa_\beta
\Omega^0_{22}(\bp)&=&\Omega^0_{21}(\bp)\Omega^0_{12}(\bp)\label{e:initial5}
\eea
where $\omega_0(\bp)=\sqrt{\bp^2+m^2}$ and $V=(2\pi)^3\delta^3(0)$ is the
large but finite spatial volume.
The hermiticity conditions on $\rho_u$ yield
\be\label{e:herm2}
{\Omega^0_{11}(\bp)}^*=\Omega^0_{22}(\bp)~,~{\Omega^0_{12}(\bp)}^*=
\Omega^0_{12}(\bp)~,~
{\Omega^0_{21}(\bp)}^*=\Omega^0_{21}(\bp)
\ee
The solution to \refpa{e:initial1}-\refpa{e:herm2}
 leading to a density matrix that satisfies
the delta function condition \refpa{e:delta} is
\bea\label{e:initialcond1}
\Omega^0_{11}(\bp)&=&\Omega^0_{22}(\bp)=\omega_0 \coth \beta \omega_0 \\
\label{e:initialcond2}
\Omega^0_{12}(\bp)&=&\Omega^0_{21}(\bp)=-\frac{\omega_0}{\sinh \beta
\omega_0}\\
N_u &=&\exp{\left[V\int_p \ln \frac{\omega_0}{\sinh \beta \omega_0}\right]}
\eea
The precise value of $N_u$ can be found using $\lim_{\beta\ra 0}
\int \calD^2 \phi_1 \rho_u(\phi_1\phistar_1,\phi_2\phi_2^*)=1$ which follows
from
\refpa{e:delta}.
The partition function $Z=\tr \hat{\rho}_u$ and the normalized density matrix\\
$\rho_0=\rho_u/\tr \hat{\rho}_u$ are now readily found as
\bea
& & Z=\exp\left[ -V\int _p 2\ln (2 \sinh(\frac{\beta \om _0}{2}))\right], \\
\lefteqn{\rho_0(\phi_1\phistar_1,\phi_2\phi_2^*)=\exp{\left[V \int_p \ln (2
\omega_0 \tanh\frac{\beta \omega_0}{2})\right]}\times} \nn\\
& & \exp{\left[-\int_p ((\phistar_1\phi_1+\phistar_2\phi_2)\omega_0\coth \beta
\omega_0
-\frac{\omega_0}{\sinh \beta
\omega_0}(\phistar_1\phi_2+\phistar_2\phi_1))\right]}
\eea .

\subsection{The time dependent ensemble}

We solve the Liouville equation for the density matrix $\rho(t)$
in an external field $\bA(t)$ with the initial condition specified
in the previous section, i.e.
\bea
i\pa_t{\hat{\rho}}(t)&=&[\hat{H}(t),\hat{\rho}(t)]\label{e:liouv}\\
\hat{\rho}(0)&=&\hat{\rho}_0.
\eea
The external field is constant in space but varies in time as
\be
\bE(t)=\left\{\begin{array}{cl}0&,t<0\\E{\bf
e}_3&,0<t<\tf\\0&,\tf<t\end{array}\right.~~
\bA(t)=\left\{\begin{array}{cl}0&,t<0\\-Et{\bf e}_3&,0<t<\tf\\-E\tf{\bf
e}_3&,\tf<t\end{array}\right.
\ee
where $E>0$ so that $eE<0$.
In the functional representation we find the equation for
 $\rho(\phi_1\phistar_1,\phi_2\phi_2^*)$:
\be
i\pa_t\rho(\phi_1\phistar_1,\phi_2\phi_2^*)=
\hat{H}(\phi_1,\phistar_1,\frac{\delta}{\delta \phi_1},\frac{\delta}{\delta
\phi_1^*})
\rho(\phi_1\phistar_1,\phi_2\phi_2^*)-
\hat{H}(\phi_2,\phi_2^*,\frac{\delta}{\delta \phi_2},\frac{\delta}{\delta
\phi_2^*})
\rho(\phi_1\phistar_1,\phi_2\phi_2^*)
\ee
Again we make a gaussian ansatz,
\be
\rho(\phi_1\phistar_1,\phi_2\phi_2^*,t)=N(t) \exp{\left[-\int  d^3x
d^3y~\phistar_i(\bx)\Omega_{ij}(\bx,\by,t)\phi_j(\by)\right]}
\ee
and perform a fourier transform as in the previos case
(with the specified external field the hamiltonian is still translationally
invariant).
Suppressing the time and momentum dependence, the equations obtained are
\bea
i\pa_t \ln N &=& V \int_p (\Omega_{11}-\Omega_{22}) \\
i\pa_t \Omega_{11}&=&{\Omega_{11}}^2-\omega^2-\Omega_{12}\Omega_{21} \\
i\pa_t \Omega_{12}&=&\Omega_{12}(\Omega_{11}-\Omega_{22}) \\
i\pa_t \Omega_{21}&=&\Omega_{21}(\Omega_{11}-\Omega_{22}) \\
i\pa_t \Omega_{22}&=&-{\Omega_{22}}^2+\omega^2-\Omega_{12}\Omega_{21}
\eea
where $\omega(\bp)=\sqrt{(\bp-e\bA)^2+m^2}$. The
density matrix $\rho$ is hermitian and the hermiticity conditions read
\be
{\Omega_{11}}^*=\Omega_{22}~,~{\Omega_{12}}^*=\Omega_{12}~,~
{\Omega_{21}}^*=\Omega_{21}
\ee
By inspection one finds that the solution to three of the equations are
\bea
\Omega_{12}&=&a(\Omega_{11}+\Omega_{22})\\
\Omega_{21}&=&b(\Omega_{11}+\Omega_{22})\\
\Omega_{22}&=&{\Omega_{11}}^*
\eea
where $a$ and $b$ are constants of motion determined by the initial conditions.
{}From
\refpa{e:initialcond1} and \refpa{e:initialcond2} we find $a=b=-1/2\cosh \beta
\omega_0$.
Eliminating $\Omega_{12},\Omega_{21}$ and $\Omega_{22}$ we find the
remaining two equations
\bea
i\pa_t \ln N &=& V \int_p (\Omega-\Omega^*) \\ \label{e:omega11}
i\pa_t \Omega&=&{\Omega}^2-\omega^2-
\frac{1}{4\cosh^2 \beta \omega_0}(\Omega+{\Omega}^*)^2
\eea
where we have set $\Omega_{11}=\Omega$. Hence the problem has been reduced to
finding $\Omega$. Now write
\be
\Omega=-i\frac{\pa_t y}{y}~~~~~y=r \exp{(i \theta \coth \beta\omega_0)} .
\ee
Inserting these expressions into \refpa{e:omega11} and rewriting the initial
condition, \refpa{e:initialcond1}, we find
\bea\label{e:polar}
\ddot{r}- \frac{\barom_0^2}{r^3}+\barom^2 r& =&0~~~,~~r(0)=1~~\dot{r}(0)=0\nn\\
r^2 \dot{\theta}&=&\barom_0~~,~~\theta(0)=0~~\dot{\theta}(0)=\barom_0
\eea
Here we have introduced dimensionless quantities
\be\begin{array}{c}
\bart=t\sqrt{|eE|},~~~~
\barbp=\bp/\sqrt{|eE|},~~~~\barm=m/\sqrt{|eE|},~~~~
\barbeta=\beta \sqrt{|eE|},\\
\Lambda=(\barp^1)^2+(\barp^2)^2+
\barm^2,~~~~
\barom=\sqrt{\Lambda+(\barp^3-\bart)^2},~~~~
\barV =V|eE|^{3/2}
\end{array}
\ee
and the overdots in \refpa{e:polar} denote $\pa/\pa \bart$.
Finally introducing the complex function\\ $z=r \exp(i\theta)$ the set of
equations
\refpa{e:polar}
can be collected  into the form
\bea\label{e:complexz}
\ddot{z}+\barom^2 z&=&0\nn\\
z(0)&=&1\nn\\
\dot{z}(0)&=&i \barom_0
\eea
One can now find a closed expression for $\Om$ in terms of special functions,
since the general solution of \refpa{e:complexz} is a linear combination of two
 parabolic cylinder functions \cite{grad}
\begin{equation}
D_{-\frac{i}{2}\Lambda -\frac{1}{2}}\left(\pm(1+i)(\bart-\barp^3)\right).
\end{equation}
The solution satisfying the initial conditions is easily written down but
 since the expression
is rather long we don't give it here.
Instead we end this section with the expression for the
time dependent normalized density matrix,
\bea\label{e:timedensity}
\lefteqn{
\rho(t)=\exp{\left[V\int_p  \ln \left( (\Omega+{\Omega}^*)\tanh \beta
\omega_0\tanh\frac{\beta \omega_0}{2}\right)\right]} \times}\nn\\
&
&\exp{\left[-\int_p~\left(\phistar_1\Omega\phi_1-
(\phistar_1\phi_2+\phistar_1\phi_2)
(\Omega+{\Omega}^*)\frac{1}{2 \cosh\beta \omega_0}
+\phistar_2{\Omega}^*\phi_2\right)
\right]} ,
\eea
where $\Om$ in terms of $r=|z|$ reads
\be\label{e:cov}
\Om(t)=\sqrt{|eE|}\left(\frac{\barom_0\coth{\barbeta \barom_0}}{r^2}
-i\frac{\dot{r}}{r}\right) .
\ee

\subsection{Expectation value of the number operator}

Once the density matrix has been found one can start asking questions about the
system.
We will be concerned with particle creation and therefore concentrate on
thermal
expectation values of the number operator,

\bea\label{e:nav}
\av{\hat N}(t)&=&\tr (\hatN \hat{\rho}(t))\nn\\
&=&\int\calD^2\phi_1\calD^2\phi_2\delta(\phi_1-\phi_2)\delta(\phi_1^*-\phi_2^*)
\hatN(\phi_1,\phistar_1)\rho(\phi_1\phistar_1,\phi_2\phi_2^*;t).
\eea
 More specifically we will calculate
the number of particles at $t=\tf$ just after the field has been turned off.
We then find ourselves in the gauge $\bA=-E\tf{\bf e}_3$. In order to make use
of
operators expressed in terms of fields in the $\bA=0$ gauge we make a
transformation to that gauge. The fields transform according to
\be
\bA(x)\ra\bA(x)-\nabla \lambda(x)~~~~~\phi(x)\ra \phi(x)e^{-ie\lambda(x)}.
\ee
Since the physical states are gauge invariant the covariance transforms as
\be
\Om(x,y)\ra\Om^g(x,y)=\Om(x,y)e^{-ie(\lambda(x)-\lambda(y))}.
\ee
For the specific transformation that changes $\bA=-E\tf{\bf e}_3$ into $\bA=0$,
the effect on the
covariance in the density matrix \refpa{e:cov} simply amounts
to a shift in the momentum
\be
\bp\ra \bp _g=\bp + e\bA(\tf)=(p^1,p^2,p^3-eE\tf)
\ee
Hence $r$,$\om$ and $\om_0$ transform as follows
\bea
r(\bp)&\ra& r_g(\bp)=r(\bp _g)\\
\barom(\bp)&\ra&\barom^g(\bp)=\sqrt{\Lambda+(\barp^3-\bart+\bartf)^2}\\
\barom_0(\bp)&\ra&\barom_0^g(\bp)=\sqrt{\Lambda+ (\barp^3+\bartf)^2}.
\eea

Now find the number operator $\hatN$. The gauge invariant stationary
groundstate,
expressed in terms of fields in the $\bA=0$ gauge,
is given by
\be
\Psi_0(\phi,\phistar)=\det (\om_0)^{-1/2} \exp{\left[-\int_p
\phistar\om_0\phi\right]}.
\ee
Associated with this groundstate are the annihilation operators
\bea
\hat{a}(\bp)&=&\frac{1}{\sqrt{2\om_0(\bp)}}(\om_0(\bp)\hat{\phi}(\bp)+
i\hat{\pi}^\dagger(\bp))\\
\hat{b}(\bp)&=&\frac{1}{\sqrt{2\om_0(\bp)}}(\om_0(\bp)
\hat{\phi}^\dagger(\bp)+i\hat{\pi}(\bp)).
\eea
The states created by $\hat a^\da (\bp )$ have charge $e$ and physical momentum
$\bp$,
while states created by $\hat b^\da (\bp )$ have charge $(-e)$ and physical
momentum
$(-\bp )$, as seen from
the momentum and charge operators expressed in terms of $a$'s and $b$'s,
\bea
\hat{\bf P}&=&-\int d^3x\; (\nabla \hat \phi \; \hat \pi +\nabla \hat \phi ^\da
\; \hat \pi ^\da ) =
\int_p \bp(\hat{a}^\dagger(\bp)\hat{a}(\bp)-\hat{b}^\dagger(\bp)\hat{b}(\bp))\\
\hat{Q}&=&ie\int d^3x\; (\hat \phi ^\da \hat \pi ^\da-\hat \phi \hat \pi )=
e\int_p(\hat{a}^\dagger(\bp)\hat{a}(\bp)-\hat{b}^\dagger(\bp)\hat{b}(\bp)).
\eea
The expression for the number operator in the $\bA=0$ gauge is
\bea
\hatN&=&\int_p(\hat{a}^\dagger(\bp)\hat{a}(\bp)+
\hat{b}^\dagger(\bp)\hat{b}(\bp))\\
&=&\int_p(\frac{1}{\om_0
(\bp)}\hat{\pi}^\dagger(\bp)\hat{\pi}(\bp)+\om_0(\bp)
\hat{\phi}^\dagger(\bp)\hat{\phi}(\bp)-V).
\eea
By \refpa{e:nav} and the gauge transformed versions of
\refpa{e:timedensity} and \refpa{e:cov}, we find after some manipulations the
result
\bea\label{e:numop}
\av{\hat N}(t_f)&=&\frac{V}{2}\int_p\left[
(\frac{\barom_0}{\barom_0^g} r_g^2 +
\frac{1}{\barom_0\barom_0^g}\dot{r}^2_g+\frac{\barom_0^g}{\barom_0} r^{-2}_g)
\coth\frac{\barbeta\barom_0^g}{2}-2\right] .
\eea
\begin{figure}[t]
\centerline{\psfig{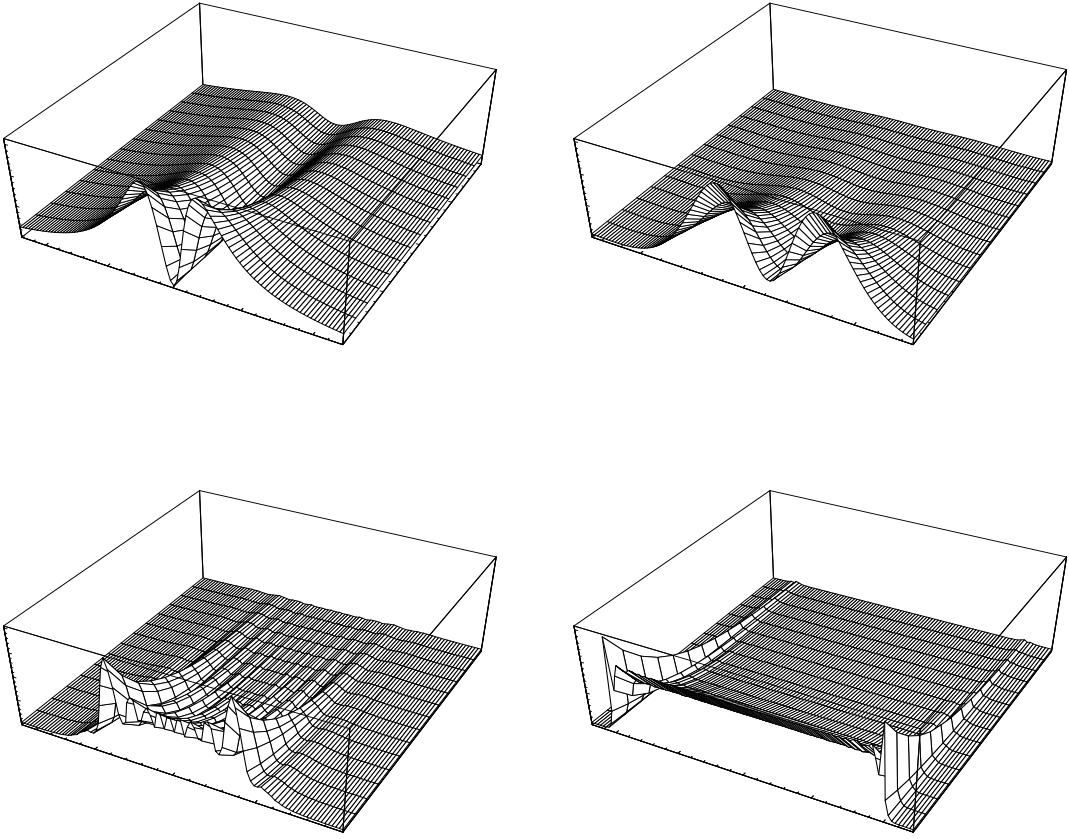}}
\caption{{\it The momentum distribution, $n(-\bp )/2$,  of the positively
charged
bosons at zero
temperature. All quantities are expressed in units of the electric
field.\label{figb1}}}
\end{figure}

\begin{figure}[t]
\centerline{\psfig{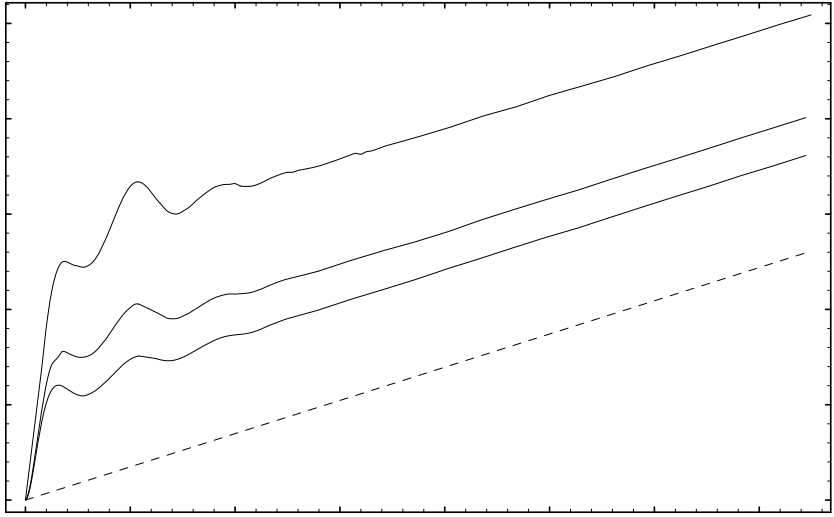}}
\vspace{1cm}
\caption{{\it The average number of created bosons for $\barm =1$.
 The dashed line shows
the asymptotic behaviour at zero temperature. All quantities are expressed in
units of the electric field.}}
\label{figb2}
\end{figure}

To find the number of particles actually created during the process
$\av{\hat N}^{\rm cr}$, we subtract the
number of particles at $t=0$:
\be
\av {\hat N} |_{t=0} =V\int_p\left[ \coth\frac{\barbeta\barom_0}{2}-1\right].
\ee
Hence, we obtain
\bea\label{e:ncr}
\av{\hat N}^{\rm cr} (t_f) &=& V \int_p n(\bp,\tf)
\coth\frac{\barbeta\barom_0^g}{2}
\nn \\
&=& \frac{\barV}{2 (2\pi )^2} \int _{-\infty} ^\infty d\barp ^3 \int _{\barm
^2} ^\infty
d\Lambda \; n(\Lambda ,\barp ^3,\bart _f) \coth\frac{\barbeta\barom_0^g}{2},
\eea
where we have defined the zero temperature number density of the created
particles
$n(\bp,\tf)$ as
\be\label{e:numberdensity1}
n=\frac{1}{2\barom_0\barom_0^g}\left( (\barom_0 r_g - \frac{\barom_0^g}{
r_g})^2
+\dot{r}^2_g\right)
\ee
Equations \refpa{e:ncr} and \refpa{e:numberdensity1} constitute
the exact expressions for the expectation value of the numbers of particles
created by the external field at finite temperature, when a time $t_f$ has
elapsed.
Since, as we have argued, the positively charged particles created by
$\hat b ^\da (\bp )$ have a momentum $(-\bp )$, their momentum distribution at
zero temperature will be given by
$n(-\bp )/2=\av{\hat b ^\da (- \bp ) \hat b (-\bp )}_{\beta =\infty}$.
Figure~\ref{figb1} shows this
distribution. Note that for small times, there are very few particles with zero
momentum. The result of integrating \refpa{e:ncr} numerically for a mass $\barm
=1$
is shown in
figure~\ref{figb2}. We see that increasing the temperature increases the number
of
created particles but for large enough $t_f$ the slope is independent of
temperature.

We have so far only considered a simple time dependence of the electromagnetic
field,
but it is straightforward to generalize to an arbitrary time dependence.
Introduce
the two dimensionless functions $f(\bart)$, $F(\bart)$ and let
\be
\bE(t)=\left\{\begin{array}{cl}0&,t<0\\Ef(\bart){\bf
e}_3&,0<t<\tf\\0&,\tf<t\end{array}\right.~~
\bA(t)=\left\{\begin{array}{cl}0&,t<0\\-EF(\bart)/\sqrt{|eE|}~~{\bf
e}_3&,0<t<\tf\\
-EF(\bartf)/\sqrt{|eE|}~~{\bf e}_3&,\tf<t\end{array}\right.\label{e:at}
\ee
The functions $f(\bart)$ and $F(\bart)$ must satisfy
\be
\dot{F}(\bart)=f(\bart)~~,~~F(0)=0
\ee
As a matter of fact, \refpa{e:ncr} and \refpa{e:numberdensity1} still hold,
with the
modifications that $r_g$ satisfies
the gauge transformed version of \refpa{e:polar}, where now
\be
\barom^g=\sqrt{\Lambda+(\barp^3-F(\bart)+F(\bartf))^2}
\ee
and similarily for $\barom_0$ and $\barom_0^g$. Because of the new time
dependence,
however,
$r_g$ cannot be expressed in terms of special functions, but has to
be found numerically. This is in practice not a serious drawback, since the
integral still must be evaluated numerically.

\subsection{Asymptotic expansions}

We have obtained the exact expression for the number of particles
created for arbitrary times $\tf$.
It is, however,  possible to obtain some more convenient expressions when
  $\tf$ is large.
To that end we investigate the behaviour of the number density $n$. The number
density may be written
\be\label{e:numberdensity2}
n(\Lambda,\barp^3,\bartf)=\frac{\barom_0^2
|z_g(\bartf)|^2+|\dot{z}_g(\bartf)|^2}
{i\barom_0(\dot{z}_g^*(\bartf) z_g(\bartf)-\dot{z}_g(\bartf) z_g^*(\bartf))}-1
\ee
where $z_g(\bart)$ by \refpa{e:complexz} satisfies
\be\label{e:complexz2}
\ddot{z}_g+(\bart-\bartf-\barp^3)^2 z_g=0.
\ee

Since the number density $n$ is not sensitive to a rescaling of $z_g$, the
relevant
initial condition for $z_g$ following from \refpa{e:complexz} is
\be\label{e:initz}
\frac{\dot{z}_g(0)}{z_g(0)}=i\barom_0^g .
\ee
Let us now be more specific and consider a large final time $\bartf>>1$.
For such $\bartf$, consider $\Lambda$ and $\barp^3$ satisfying
\begin{equation}\label{e:int}
-\bartf-\sqrt{\Lambda}<<\barp^3<<-\sqrt{\Lambda}~~,
{}~~\barp^3<<-1~~,~~\bartf>>\sqrt{\Lambda}.
\end{equation}
For $\Lambda$ and $\barp^3$ restricted by \refpa{e:int} the initial condition
\refpa{e:initz} reads
\be\label{e:initz2}
\frac{\dot{z}_g(0)}{z_g(0)}\sim i(\bartf+\barp^3)
\ee
A solution to \refpa{e:complexz2} satisfying \refpa{e:initz2} is
\be
z_g(\bart)=D_{-\frac{i}{2}\Lambda
-\frac{1}{2}}\left(-(1+i)(\bart-\bartf-\barp^3)\right),
\ee
which may be confirmed using the first of the following
asymptotic expansions of parabolic cylinder functions
when $|x|>>1,|x|>>|\nu|$ \cite{grad}:
\bea\label{e:asymp1}
D_\nu(x e^\frac{i\pi}{4})&\sim&e^{-\frac{i x^2}{4}}
 (x e^\frac{i\pi}{4})^\nu~~~,~~x>0\\
D_\nu(x e^\frac{i\pi}{4})&\sim&e^{-\frac{ix^2}{4}}
 (x e^\frac{i\pi}{4})^\nu -\frac{\sqrt{2\pi}}{\Gamma(-\nu)}
e^{\frac{ix^2}{4}}e^{-i\pi\nu} (x e^\frac{i\pi}{4})^{-\nu-1}~~~,~~x<0.
\label{e:asymp2}
\eea
It is now a straightforward calculation to find the asymptotic behaviour of
 $z_g(\bartf)$ in which case \refpa{e:asymp2} applies.
By \refpa{e:numberdensity2}, and using
$\barom_0\sim|\barp^3|$ and $|\Gamma(\frac{i}{2}\Lambda+\frac{1}{2})|^{-2}=
\frac{1}{\pi}\cosh \frac{\pi\Lambda}{2}$,
one finds the number density
\be
n(\Lambda,\barp^3,\bartf)\sim 2 e^{-\pi \Lambda}.
\ee
What happens if $\barp$ lies outside the region in \refpa{e:int}?
It is clear that for $|\barp ^3|$ sufficiently large compared to
 $\bartf$, we can neglect the time dependence in the equation for $z_g$.
 Hence there will be no created particles i.e. $n(\Lambda ,\barp ^3,\bart
_f)=0$. In
the region where $| \bp |$  is neither large, nor restricted by \refpa{e:int},
the
contribution to $\av{\hat N}^{\rm cr}$ will be small compared to the
dominant contribution. Therefore we can effectively set $n$ to zero in this
region.
We conclude that for $\bartf >>1$  and $\bartf >> \barm$,
\begin{equation}\label{e:numberdensity3}
n(\Lambda ,\barp ^3,\bartf)=\left\{ \begin{array}{ll}
2 e^{-\pi \Lambda}, \; \; \; & -\bart _f < \barp ^3 < 0, \\
0, \; \; \;  & \mbox{otherwise} .
\end{array} \right.
\end{equation}
and thus
\be
\av{\hat N}^{\rm cr}(\bart _f)=\frac{\barV}{2(2\pi)^2}\int_{-\tf}^{0}d\barp^3
\int_{\barm^2}^{\infty} d\Lambda~~2 e^{-\pi \Lambda}
\coth\frac{\barbeta\barom_0^g}{2}.
\ee
Two limits involving $\beta$ may now be considered, namely $1/\barbeta<<\bart
_f$
and $1/\barbeta>>\bartf$,
\bea\label{ncrzero}
\av{\hat N}^{\rm cr}(\bart _f) &=&\frac{\barV}{4\pi^3}\bartf e^{-\pi\barm^2}=
\frac{ \alpha }{\pi^2 }VE^2 t_f  e^{-\frac{\pi m^2}{|e E|}}~~,~~
\frac{1}{\barbeta}<<\bart _f\\ \label{e:limit2}
\av{\hat N}^{\rm cr}(\bart _f) &=&\frac{\barV}{2\barbeta\pi^3}\ln(2\bartf)
e^{-\pi\barm^2}
=(\frac{\alpha }{\pi})^{1/2}\frac{V|E|}{\beta\pi^2}\ln\left(2\tf\sqrt{|e
E|}\right)
e^{-\frac{\pi m^2}{|e E|}},~~\frac{1}{\barbeta}>>\bartf
\eea
The result \refpa{ncrzero} is identical to the zero temperature result
obtained in \cite{fradkin}, eq. (5.1.61).

\subsection{Probabilities}

The probability $P(\Psi)$ of finding the system in a normalized state
$\ket{\Psi}$ is
\be
P(\Psi)=\bra{\Psi}\hat{\rho}\ket{\Psi}
\ee
We are interested in the probability $P(n,m;\tf)$  of finding the system
in a state with $n$ particles and $m$ anti particles at time $t=\tf$.
For specific momenta these states are
\be
\ket{\bp_1...\bp_n,\bk_1...\bk_m}=\frac{1}{V^{(n+m)/2}}\frac{1}{\sqrt{n! m!}}
\prod_{i=1}^{n} \prod_{j=1}^{m}
\hat{a}^\dagger(\bp_i)\hat{b}^\dagger(\bk_j)\ket{0}.
\ee
The desired probability is a sum of probabilities for such states with
$n$ and $m$ held fixed, i.e.
\be
P(n,m;\tf)=V^{n+m}\int_{p_1...p_n}\int_{k_1...k_m}
\bra{\bp_1...\bp_n,\bk_1...\bk_m}\hat{\rho}(\tf)
\ket{\bp_1...\bp_n,\bk_1...\bk_m}
\ee
One finds that the probabilities, $P(n,m;\tf)$, are generated by a
function $f(x,y;\tf)$ as
\be\label{e:prob}
P(n,m;\tf)= \left.f(1,1;\tf)^{-1}\frac{1}{n! m!}\frac{\pa^{n+m}}{\pa x^n\pa
y^m}f(x,y;\tf)\right|_{x=y=0}
\ee
It immediately follows that $\sum_{n,m=0}^{\infty}P(n,m;\tf)=1$.
The generating function $f(x,y;\tf)$ reads
\be
f(x,y;\tf)=\exp{\left[-\Tr \ln \left\{ (1-x g(\tf))(1-y g(\tf))-x y h(\tf)
\right\}
\right]}
\ee
where the functions $g(\tf)$ and $h(\tf)$ are expressed in terms of the
number density $n(\tf)$
defined in \refpa{e:numberdensity1}. Suppressing the momentum dependence, we
have
\bea
g(\tf)&=&\frac{1}{(n(\tf)+1+\coth \barbeta \barom_0^g)\sinh\barbeta
\barom_0^g}\\
h(\tf)&=&\frac{n(\tf)(n(\tf)+2)}{(n(\tf)+1+\coth\barbeta \barom_0^g)^2}
\eea
We note that when $\barbeta=\infty$ only $h(\tf)$ survives. Since $h(\tf)$
multiplies
$x y$ we conclude that only uncharged states are present at zero temperature.
This
is what we expect since all states have been created out of the uncharged
groundstate.
On the other hand when $\tf=0$ we have that $h(0)=0$ while
 $g(0)=\exp(-\beta \om_0)$, leading to ordinary Boltzmann probabilities
for the initial ensemble.

The probability $P(0,0;\tf)$ is readily obtained
\bea
P(0,0;\tf)&=&f(1,1;\tf)^{-1}\nn \\
&=&\exp\left[- \Tr \ln \left\{\frac{1}{2} \coth(\frac{\barbeta \barom_0^g}{2})
\left(n(\tf)+1+\coth\barbeta \barom_0^g\right)\right\}\right]
\label{pzero}
\eea
At zero temperature this reduces to
\be
P(0,0;\tf)=\exp\left[-\Tr \ln(1+\frac{n(\tf)}{2})\right]~~,~~\barbeta=\infty
\ee
which for large times, $\bartf>>1$, by \refpa{e:numberdensity3} becomes
\be
P(0,0;\tf)=\exp\left[ -\frac{\barV\bartf}{2(2\pi)^2}
\int_{\barm^2}^{\infty} d\Lambda \ln (1+e^{-\pi \Lambda})\right]
\ee
This is the result obtained by Schwinger in 1951 \cite{schwinger}.
We end this section by noting that expectation values of some operators may be
obtained from the generating function. For the number operator we have
\bea
\av{\hatN}(\tf)&=&\sum_{n,m=0}^{\infty}(n+m)P(n,m;\tf)\nn \\
&=&f(1,1;\tf)^{-1}\left.\pa_x f(x,x;\tf)\right|_{x=1} \nn \\
&=&\Tr \lbrack (n(\tf)+1)\coth\frac{\barbeta \barom_0^g}{2}-1\rbrack ,
\eea
which of course agrees with \refpa{e:numop}.

\section{Fermions}
We will set up the formalism by studying an illustrative example, the
two-dimensional fermionic harmonic oscillator.
The annihilation and creation operators, $\hat a, \hat b$ and $\hat a^\da ,
\hat b
^\da $, satisfy
\begin{equation}
\{ \hat a,\hat a^ \da \}=\{ \hat b,\hat b^ \da \}=1.
\end{equation}

All other anticommutators vanish. The Hilbert space is spanned by the four
orthogonal
states $\ket{0,0}, \ket{1,0}, \ket{0,1}$ and $\ket{1,1}$,
\begin{equation}
\hat a \ket{0,0}=\hat b \ket{0,0}=0, \; \; \hat a^\da \ket{0,0}=\ket{1,0} , \;
\;
\hat b^\da \ket{0,0}=\ket{0,1}, \; \; \hat a^\da \hat b^\da \ket{0,0}=\ket{1,1}
{}.
\end{equation}
The states are normalized, $\amp{0,0}{0,0}=\amp{1,0}{1,0}=\amp{0,1}{0,1}=
\amp{1,1}{1,1} =1$. We represent this system by functions $\Psi (\eta ^*,\eta)$
of
a complex Grassmann number $\eta$ and its complex conjugate $\eta ^*$ in the
following
manner,
\begin{equation}
\begin{array}{cccc}
\ket{0,0} \lr 1, & \ket{1,0} \lr \eta ^* , & \ket{0,1} \lr \eta ,
& \ket{1,1} \lr \eta ^* \eta , \\
\hat a \lr \pa _{\eta ^*}, & \hat a ^\da \lr \eta ^* ,
& \hat b \lr \pa _\eta , & \hat b ^\da \lr \eta ,
\end{array}
\end{equation}
so that overlaps with Grassmann states are given by
\begin{equation}
\amp{\eta ^* \eta}{0,0}=1, \; \; \amp{\eta ^* \eta }{1,0}=\eta ^*, \; \;
\amp{\eta ^* \eta }{0,1}=\eta , \; \; \amp{\eta ^* \eta}{1,1}=\eta ^* \eta
\end{equation}
and
\bea
\amp{\eta ^*,\eta}{{\eta '}^*,\eta '}&=&\sum_{n,m=0}^1 \amp{\eta ^*,\eta}{n,m}
\amp{n,m}{{\eta '}^*,\eta '}\\
& =&1+\eta ^* \eta '-{\eta '}^* \eta +\eta ^* \eta {\eta '}
^* \eta '=e^{\eta ^* \eta '-{\eta '}^* \eta} .
\eea
The Grassmann state may now be expanded as
\begin{equation}
\ket{\eta ^* \eta}=\sum _{n,m=0}^1 \ket{n,m} \amp{n,m}{\eta ^* \eta}=
\ket{0,0}+\ket{1,0} \eta +\ket{0,1} \eta ^* +\ket{1,1} \eta ^* \eta .
\end{equation}
The partition of unity can thus be expressed as
\bea
\hat 1&=&\ket{0,0} \bra{0,0}+\ket{1,0} \bra{1,0}+\ket{0,1} \bra {0,1}+\ket{1,1}
\bra{1,1}\nn \\
& =&\int d^2 \eta ' d^2 \eta \ket{\eta ^* \eta } \amp{\eta ^* \eta}
{{\eta '}^* \eta '} \bra{{\eta '}^* \eta '},
\eea
where $d^2 \eta =d\eta ^* d \eta$. Hence an arbitrary overlap is given by
\begin{eqnarray}
\amp{\Psi _1 }{\Psi _2} &=& \int d^2 \eta ' d^2 \eta \amp{\Psi _1} {\eta ^*
\eta}
\amp{\eta ^* \eta}{{\eta '}^* \eta '} \amp{{\eta '}^* \eta '}{\Psi _2} \nn \\
&=& \int d^2 \eta ' d^2 \eta \; e^{\eta ^* \eta '-{\eta '}^* \eta} \;
\Psi _1 ^* (\eta ^* , \eta) \Psi _2 ({\eta '}^* , \eta ') .
\end{eqnarray}
Let $\hat {\cal O}$ be an operator,
\begin{equation}
\hat {\cal O}=\sum _{n_1,n_2,m_1,m_2=0}^1 {\cal O} _{n_1 n_2, m_1 m_2}
\ket{n_1,n_2} \bra{m_1,m_2} .
\end{equation}
The trace of this operator is then,
\begin{equation}
\tr \hat {\cal O} ={\cal O} _{00,00}+{\cal O} _{10,10}+{\cal O} _{01,01} +
{\cal O} _{11,11} .
\end{equation}
This can be expressed as,
\begin{equation}
\tr \hat {\cal O} =\int d^2 \eta ' d^2 \eta  \bra{\eta ^* \eta} \hat {\cal O}
\ket{{\eta '}^* \eta '} \amp{\eta ^* \eta}{{\eta '}^* \eta '} .
\end{equation}
The ``naive'' trace,
\begin{equation}
\int d^2 \eta ' d^2 \eta  \bra{{\eta '}^* \eta '} \hat {\cal O}
\ket{\eta ^* \eta } \amp{\eta ^* \eta}{{\eta '}^* \eta '}=
{\cal O} _{00,00}-{\cal O} _{10,10}-{\cal O} _{01,01}+{\cal O} _{11,11},
\end{equation}
is in fact the supertrace $\mbox{Str} (\hat {\cal O})=\mbox{tr} (\hat {\cal O}
(-1) ^{\hat F})$. Let us now
 turn to field theory. The Hamiltonian for Dirac fermions in an external
electromagnetic field ${\bf A}$ is,
\begin{equation}
\hat H =\int d^3x d^3y \frac{1}{2}\hat \lbrack \psi ^\da _\alpha (\bx )
,\hat \psi _\beta (\by ) \rbrack h_{\alpha \beta} (\bx  ,\by  ) ,
\end{equation}
where
\begin{equation}
h(\bx  ,\by  )=(-i \gamma ^0 \bfga \cdot (\nabla _x -ie {\bf A} (\bx ))+m
\gamma ^0)
\delta (\bx -\by ).
\end{equation}
The anticommutation relations read
\begin{equation}
\{ \hp _\alpha (\bx ),\hpd _\beta (\by ) \}=\delta _{\alpha \beta} \delta (\bx
-\by ).
\end{equation}
All other anticommutators vanish. As was first done in \cite{jackiw}, we let
the
fermionic field operators act on
wavefunctionals $\Psi (\eta ^* , \eta )$ of a complex Grassmann field $\eta$
and
its complex conjugate $\eta ^*$ according to
\begin{eqnarray}
\hp _\alpha (\bx ) & \lr & \frac{1}{\sqrt{2}}(\eta _\alpha (\bx )+\fd{\eta
_\alpha ^*(\bx )}),
\label{eq:psi}\\
\hpd _\alpha (\bx ) & \lr & \frac{1}{\sqrt{2}}(\eta ^* _\alpha (\bx )+\fd{\eta
_\alpha (\bx )}).
\label{eq:psid}\end{eqnarray}
The representation defined by \refpa{eq:psi} and \refpa{eq:psid} of the field
operator
algebra is reducible. This can be seen explicitly by introducing the operators,
\begin{eqnarray}
\hat \theta _\alpha (\bx ) & \lr & \frac{1}{\sqrt{2}}(\eta _\alpha (\bx
)-\fd{\eta _\alpha ^*(\bx )}),
\nn \\
\hat \theta ^\da _\alpha (\bx ) & \lr & \frac{1}{\sqrt{2}}(-\eta ^* _\alpha
(\bx )+\fd{\eta _\alpha (\bx )}).
\end{eqnarray}
We have
\begin{equation}
\{ \hat \theta _\alpha (\bx ),\hat \theta ^\da  _\beta (\by ) \}=\delta
_{\alpha \beta} \delta (\bx -\by ),
\end{equation}
and $\hat \theta $ as well as $\hat \theta ^\da$ anticommute with both $\hp$
and $\hpd$ making the
representation reducible.
Contrary to what is said in \cite{jackiw} we may now view wavefunctionals as
overlaps
with Grassmann field states $\ket{\eta ^* \eta}$,
\begin{eqnarray}
\amp{\eta ^* \eta}{{\eta '} ^* \eta '} &=& \exp (\int d^3x (\eta ^*_\alpha (\bx
) \eta '
_\alpha (\bx )-{\eta '}^*_\alpha (\bx ) \eta _\alpha (\bx ))), \nn \\
\Psi (\eta ^* , \eta ) &=& \amp{\eta ^* \eta}{\Psi}.
\end{eqnarray}
The partition of unity is given by functional Grassmann integration,
\begin{equation}
\hat 1=\int D^2 \eta ' D^2 \eta \ket{\eta ^* \eta}\amp{\eta ^* \eta}
{{\eta '} ^* \eta '}\bra{{\eta '} ^* \eta '},
\end{equation}
having set $D^2 \eta=D \eta ^* D\eta$ and hence the inner product is
\begin{equation}
\amp{\Psi _1}{\Psi _2}=\int D^2 \eta ' D^2 \eta \; \amp{\eta ^* \eta}
{{\eta '} ^* \eta '} \; \Psi_1 ^*(\eta ^*,\eta) \Psi _2({\eta '}^*,\eta ').
\end{equation}
For an operator $\hat {\cal O}$ we have
\begin{eqnarray}
\bra{\eta ^* \eta}\hat {\cal O} \ket{{\eta '} ^* \eta '} &=&
 {\cal O}(\eta ^* \eta,{ \eta '} ^* \eta ' )\nn \\
\bra{\eta ^* \eta}\hat {\cal O} (\hp , \hpd)\ket{\Psi} &=& \hat {\cal O}
(\eta ^*,\fd{\eta ^*},\eta, \fd{\eta}) \amp{\eta ^* \eta}{\Psi},\nn \\
\mbox{tr}_d(\hat {\cal O}) &=& \int D^2 \eta ' D^2 \eta \bra{\eta ^* \eta}\hat
{\cal O}
 \ket{{\eta '} ^* \eta '} \amp{ \eta ^* \eta}{{\eta '} ^* \eta '}.
\end{eqnarray}
Since the representation is reducible, all states will have an unphysical
degeneracy
with the degree of degeneracy given by $d$. Hence the physical trace is given
by
\begin{equation}
\mbox{tr}(\hat {\cal O})=\frac{1}{d} \mbox{tr}_d (\hat {\cal O}).
\end{equation}
The degeneracy $d$ is given by the number of states generated by
$\hat \theta ^\da$ and $\hat \theta$.
For symmetry reasons this is equal to the number of states generated by $\hpd$
and
$\hp$. The total number of states in the functional space is $\mbox{tr}_d(\hat
1)$.
Hence,
\begin{equation}
d^2=\mbox{tr}_d(\hat 1)=\int D^2 \eta ' D^2 \eta \amp{\eta ^* \eta}
{{\eta '} ^* \eta '} ^2=\det(4I)=
\exp(4V\int _p \ln 4),
\end{equation}
where $I_{\alpha \beta}(\bx ,\by )=\delta _{\alpha \beta} \delta (\bx -\by )$.
Thus
\begin{equation}
d=\det (2I)=\exp (4V \int _p \ln 2).
\end{equation}
Let us end this subsection by giving a standard formula for Gaussian functional
integrals,
\begin{eqnarray}
\lefteqn{\int D^2\eta \exp (\int d^3x d^3y \; \eta ^*_\alpha (\bx ) \Omega
_{\alpha \beta}(\bx  ,\by  )
\eta _\beta (\by )+\int d^3x \; \eta ^*_\alpha (\bx )\chi _\alpha (\bx )+
\xi _\alpha ^* (\bx )
\eta _\alpha (\bx ))=}\nn \\
&& \det (-\Omega) \exp (-\int d^3x d^3y \; \xi _\alpha ^* (\bx )
\Omega _{\alpha \beta}^{-1}(\bx  ,\by  ) \chi _\beta (\by )) \hspace{6cm}
\end{eqnarray}
\subsection{Initial ensemble}
We will now do all the calculations for fermions that were previously done
for bosons. Let $\hat H_0$ denote
the Hamiltonian with ${\bf A}=0$. By \refpa{e:ini1} and \refpa{e:ini2} we find,
letting \\
$\rho _u(\eta _1^* \eta _1,\eta _2^*
\eta _2)=\bra{\eta _1^*,\eta _1}\hat \rho _u\ket{\eta _2 ^*,\eta _2}$,
\begin{eqnarray}
-\pad{\beta}\rho _u(\eta _1^* \eta _1,\eta _2^* \eta _2) &=&\hat
H_0 (\eta _1^*, \fd{\eta _1^*},\eta _1, \fd{\eta _1})
\rho _u(\eta _1^* \eta _1,\eta _2^*\eta _2),\label{eq:h0r}\\
\lim _{\beta \rightarrow 0} \rho _u(\eta _1^* \eta _1,\eta _2^* \eta _2) &=&
\amp{\eta _1^*,\eta _1}{\eta _2 ^*,\eta _2},\label{eq:beta0}\\
\rho _u(\eta _1^* \eta _1,\eta _2^* \eta _2) &=&
\rho _u ^*(\eta _2^* \eta _2,\eta _1^* \eta _1)\label{eq:herm}.
\end{eqnarray}
Make a Gaussian ansatz (suppressing spinor indices),
\begin{equation}
\rho _u(\eta _1^* \eta _1,\eta _2^* \eta _2) =N_u \exp (\int d^3x d^3y \;
\eta ^*_i (\bx ) \Omega ^0 _{ij} (\bx ,\by )\eta _j (\by )) \; \; \; \;
i,j=1,2.
\end{equation}
Using \refpa{eq:h0r} we obtain
\begin{eqnarray}
-\pa _\beta \ln N_u &=& \frac{1}{2} \mbox{Tr} (h^0 \Om
^0_{11}),\label{eq:first} \\
-\pa _\beta \Om ^0_{11} &=& \frac{1}{2}(I-\Om ^0 _{11}) h^0 (I+\Om ^0 _{11}),\\
-\pa _\beta \Om ^0 _{12} &=& \frac{1}{2}(I-\Om ^0 _{11})h^0 \Om ^0 _{12},\\
-\pa _\beta \Om ^0 _{21} &=& -\frac{1}{2}\Om ^0 _{21} h^0 (I+\Om ^0 _{11}),\\
-\pa _\beta \Om ^0 _{22} &=& -\frac{1}{2} \Om ^0 _{21} h^0 \Om ^0 _{12} ,
\end{eqnarray}
where $h^0$ is $h$ having set ${\bf A}=0$. Furthermore, by \refpa{eq:beta0},
\begin{eqnarray}
\lim _{\beta \rightarrow 0} N_u &=& 1,\\
\lim _{\beta \rightarrow 0} \Om ^0 _{11} &=& \lim _{\beta \rightarrow 0} \Om ^0
_{22}=0,\\
\lim _{\beta \rightarrow 0} \Om _{12}^0 &=& -\lim _{\beta \rightarrow 0} \Om
^0_{21}=I,
\end{eqnarray}
and by \refpa{eq:herm},
\begin{equation}
(\Om ^0 _{11} )^\da =\Om ^0 _{22}, \; \; \;  (\Om ^0 _{12} )^\da =\Om ^0 _{12},
\; \;
 \; (\Om ^0 _{21} )^\da =\Om ^0 _{21},\label{eq:last}
\end{equation}
where $(\Om ^\da )_{\alpha \beta}(\bx ,\by )=\Om ^* _{\beta \alpha}
(\by ,\bx )$. Fourier transforming, just like for bosons,
\begin{equation}
\Om (\bx ,\by )=\int _p \Om (\bp )e^{i \bp \cdot (\bx -\by )}, \; \; \;
\eta (\bx )=\int _p \eta (\bp )e^{i \bp \cdot \bx},
\end{equation}
we find the solution satisfying \refpa{eq:first}-\refpa{eq:last},
\begin{eqnarray}
\Om _{11}^0 (\bp ) &=& \Om ^0_{22}(\bp )=-\gamma ^0 \frac{(\bfga \cdot
\bp+m)}{\om _0}
\tanh (\frac{\beta \om _0}{2}),\label{eq:om11sol}\\
\Om ^0 _{12}(\bp ) &=& \Om _{11}(\bp )+1,\; \; \; \Om ^0_{21}(\bp )=
\Om ^0 _{11} (\bp )-1,\\
N_u &=& \exp \left\lbrack 4V \int _p \ln \cosh \frac{\beta \om _0}{2}
\right\rbrack .
\end{eqnarray}
Hence one may obtain the partition function,
\begin{equation}
Z=\mbox{tr}\hat \rho _u=\exp \left\lbrack 4V \int _p \ln 2 \cosh
\frac{\beta \om _0}{2} \right\rbrack ,
\end{equation}
and the normalized density matrix, $\hat \rho _0=\hat \rho _u/\mbox{tr}\hat
\rho _u$,
\begin{eqnarray}
\lefteqn{\rho _0 (\eta _1^* \eta _1,\eta _2 ^* \eta _2)=\exp ( -4V
\int _p \ln 2 ) \times }\nn \\
 && \exp \left\lbrack \int _p ((\eta _1^*+\eta _2^*) \Om _{11}^0
 (\eta _1 +\eta _2 )+\eta _1^* \eta _2 -
\eta ^* _2 \eta _1) \right\rbrack ,
\end{eqnarray}
having suppressed the momentum dependence.
\subsection{The time dependent ensemble}
Let us now find $\hat \rho (t)$ with the initial condition
$\hat \rho (0)=\hat \rho _0$. By \refpa{e:liouv}, one has
\begin{equation}
i\pa _t\rho (\eta _1^* \eta _1,\eta _2 ^* \eta _2)=\hat H(\eta _1^*, \fd{\eta
_1^*},\eta _1, \fd{\eta _1}) \rho (\eta _1^* \eta _1,\eta _2 ^* \eta _2)-
(\hat H(\eta _2^*, \fd{\eta _2^*},\eta _2, \fd{\eta _2}) \rho (\eta _2^* \eta
_2,\eta _1 ^* \eta _1))^*.
\end{equation}
We make an ansatz of the same form as the initial condition,
\begin{eqnarray}
\lefteqn{\rho  (\eta _1^* \eta _1,\eta _2 ^* \eta _2,t)=\exp ( -4V
\int _p \ln 2 ) \times }\nn \\
 && \exp \left\lbrack \int _p ((\eta _1^*+\eta _2^*) \tanh (\frac{\beta \om
_0}{2}) \; \Om(t)
 \; (\eta _1 +\eta _2 )+\eta _1^* \eta _2 -
\eta ^* _2 \eta _1) \right\rbrack ,
\end{eqnarray}
in momentum space. Since $\hat \rho$ is hermitian, $\Om (\bp ,t)$ is a
hermitian matrix.
Hence, with
the same external field as for bosons, using \refpa{eq:om11sol} we obtain
\begin{equation}
i \pa _t \Om =\lbrack h,\Om \rbrack, \; \; \; \Om (\bp ,0)=-
\gamma ^0 \frac{(\bfga \cdot \bp+m)}{\om _0 (\bp )}\label{eq:omt}.
\end{equation}
Introducing the spinors $\chi _1$ and $\chi _2$,
\begin{equation}
1+\Om =\chi _1 \chi _1 ^\da+\chi _2 \chi _2 ^\da ,
\end{equation}
where,
\begin{equation}
i \pa _t \chi _j=h \chi _j, \; \; \; j=1,2 ,\label{eq:chi}
\end{equation}
we solve \refpa{eq:omt}. Noting that,
\[ h(\bp )=\gamma ^0 (\bfga \cdot (\bp -e {\bf A})+m), \]
and hence $h(\bp )^2=\om (\bp )^2=(\bp -e {\bf A})^2+m^2$, we differentiate
 \refpa{eq:chi} with respect to t and obtain,
\begin{equation}
\frac{\pa ^2 \chi _j}{\pa t^2}+(\om ^2+i e E \gamma ^0 \gamma ^3) \chi _j=0.
\label{eq:chi2}
\end{equation}
To have \refpa{eq:chi2} diagonal we choose the chiral representation of the
gamma
matrices,
\begin{equation}
\gamma ^0=\left\lbrack \begin{array}{cc} 0 & -1 \\ -1 & 0 \end{array}
\right\rbrack ,
\; \; \bfga =\left\lbrack \begin{array}{cc} 0 & \bfsi \\ -\bfsi & 0 \end{array}
\right\rbrack , \; \; \gamma ^0 \gamma ^3=\left\lbrack \begin{array}{cc}
\sigma ^3 & 0\\ 0 & -\sigma ^3 \end{array} \right\rbrack ,
\end{equation}
in which,
\begin{equation}
h(\bp )=\left\lbrack \begin{array}{cc} \bfsi \cdot (\bp -e {\bf A}) & -m \\
-m & -\bfsi \cdot (\bp -e {\bf A}) \end{array} \right\rbrack .
\end{equation}
It remains to choose initial conditions for $\chi _1$ and
$\chi _2$ such that the initial condition for $\Om$ is satisfied. This is most
easily
done by diagonalizing $\Om(t=0)$. Let
\begin{equation}
\Om (0)=U D U^\da, \; \; \; \chi _j (0)=U \chi _j^D,
\end{equation}
for some unitary matrix $U$ chosen in such a way that
$D=\mbox{diag}(1,1,-1,-1)$. Explicitly:
\begin{equation}
U=\left\lbrack \begin{array}{cccc}
\frac{m}{r_+}& \frac{-p^1+i p^2}{r_+}  & \frac{m}{r_-} & \frac{-p^1+i
p^2}{r_-}\\
 0 & \frac{p^3+\om _0}{r_+} &0 & \frac{p^3-\om _0}{r_-}\\
\frac{p^3+\om _0}{r_+} & 0 &\frac{p^3-\om _0}{r_-} & 0 \\
\frac{p^1+i p^2}{r_+} & \frac{m}{r_+} &\frac{p^1+i p^2}{r_-} & \frac{m}{r_-}
\end{array} \right\rbrack
\end{equation}
where $r_+=\sqrt{2\om _0 (\om _0+p^3)}$ and $r_-=\sqrt{2\om _0 (\om _0-p^3)}$.
Thus,
\begin{equation}
\chi _1^D (\chi _1^D)^\da +\chi _2 ^D (\chi _2 ^D)^\da
=1+D=\mbox{diag}(2,2,0,0),
\end{equation}
and we choose,
\begin{equation}
\chi _1^D=\sqrt{2} \left\lbrack \begin{array}{c} 1\\ 0 \\ 0 \\ 0\\ \end{array}
\right\rbrack , \; \; \; \; \chi _2^D =\sqrt{2} \left\lbrack \begin{array}{c}
0\\ 1 \\ 0 \\ 0\\ \end{array} \right\rbrack .
\end{equation}
leading to,
\begin{eqnarray}
\chi _1 (0) &=& \frac{1}{\sqrt{\om _0(\om _0+p^3)}}\left\lbrack
\begin{array}{c}
m\\ 0\\ p^3+\om _0\\ p^1+i p^2 \end{array} \right\rbrack ,\label{eq:c10}\\
\chi _2 (0) &=& \frac{1}{\sqrt{\om _0(\om _0+p^3)}}\left\lbrack
\begin{array}{c}
-p^1+i p^2\\p^3+\om _0\\ 0\\ m\end{array} \right\rbrack \label{eq:c20}.
\end{eqnarray}
Now $h^0$ is proportional to $\Om (0)$, $h^0=-\om _0 \Om (0)$. This implies
that
both $\chi _1(0)$ and $\chi _2(0)$ are eigenspinors of $h^0$ with the same
eigenvalue $(-\om _0)$. By \refpa{eq:chi} we then have,
\begin{equation}
\pa _t \chi _j (0)=i \om _0 \chi _j (0) .
\end{equation}
Hence we obtain, using \refpa{eq:chi2},
\begin{eqnarray}
\chi _1 &=& \frac{1}{\sqrt{\barom _0 (\barom _0+\barp ^3)}} \left\lbrack
\begin{array}{c} \barm y_1\\ 0\\ (\barp ^3+\barom _0) y_2\\(\barp ^1+i \barp
^2) y_1
\end{array} \right\rbrack ,\label{eq:chi1t}\\
\chi _2 &=& \frac{1}{\sqrt{\barom _0 (\barom _0+\barp ^3)}} \left\lbrack
\begin{array}{c} (-\barp ^1+i \barp ^2) y_1\\ (\barp ^3+\barom _0) y_2\\0\\
\barm y_1
\end{array} \right\rbrack \label{eq:chi2t},
\end{eqnarray}
where,
\begin{eqnarray}
\ddot{y_1}+(\barom ^2-i) y_1 &=&0, \; \; y_1(0)=1, \; \; \dot{y_1}(0)=i\barom
_0 ,\\
\ddot{y_2}+(\barom ^2+i) y_2 &=&0, \; \; y_2(0)=1, \; \; \dot{y_2}(0)=i\barom
_0 ,
\end{eqnarray}
having introduced dimensionless quantities as for bosons. $y_1$ can be written
as a
linear combination of the (linearly dependent) parabolic cylinder functions,
\begin{equation}
D_{-\frac{i}{2}\Lambda -1}(-(1+i) \tau),\; \;
D_{-\frac{i}{2}\Lambda -1}((1+i) \tau),\; \;
D_{\frac{i}{2}\Lambda}((1-i) \tau),\; \;
D_{\frac{i}{2}\Lambda }(-(1-i) \tau),
\end{equation}
whereas $y_2$ can be written in terms of,
\begin{equation}
D_{-\frac{i}{2}\Lambda}(-(1+i) \tau),\; \;
D_{-\frac{i}{2}\Lambda}((1+i) \tau),\; \;
D_{\frac{i}{2}\Lambda -1}((1-i) \tau),\; \;
D_{\frac{i}{2}\Lambda -1}(-(1-i) \tau) .
\end{equation}
Note that,
\begin{equation}
y_2=\frac{(\barp ^3-\bart) y_1-i \dot y_1}{\barom _0+\barp ^3} .
\end{equation}
\subsection{Expectation value of the number operator}
The number operator, expressed in terms of fields in the ${\bf A}=0$ gauge, is
simply,
\begin{equation}
\hat N=\int _p (\frac{1}{2}\lbrack \hpd _\alpha (\bp ),\hp _\beta (\bp )\rbrack
\frac{h^0_{\alpha \beta}(\bp )}{\om _0 (\bp)}+2V).
\end{equation}
As for bosons, we transform the covariance to the gauge ${\bf A}=0$ by shifting
the
momentum,
\begin{equation}
\Om ^g(\bp ,t_f)=\Om (\bp +e {\bf A}(t_f),t_f) .
\end{equation}
The expectation value of the number operator then follows,
\begin{equation}
\av {\hat N} (t_f)=\mbox{tr}(\hat N \hat \rho (t_f) )=\frac{V}{2}\int _p \left(
4-
\tanh (\frac{
\beta \om _0^g (\bp )}{2})
 \mbox{tr}\left( \Om (\bp ,0) \Om ^g (\bp ,t_f) \right) \right) .
\label{eq:exnop}
\end{equation}
Subtracting the number of particles present at $t=0$,
\begin{equation}
\av {\hat N} |_{t=0} =2V \int _p (1-\tanh (\frac{\beta \om _0}{2})),
\end{equation}
the number of particles created is,
\begin{equation}
\av {\hat N} ^{\rm cr} (t _f)=\frac{V}{2}\int _p \tanh (\frac{\beta \om _0^g
}{2})
(4-\mbox{tr}(\Om ^0 \Om ^g)) .
\end{equation}
\vspace{2mm}
\begin{figure}[t]
\centerline{\psfig{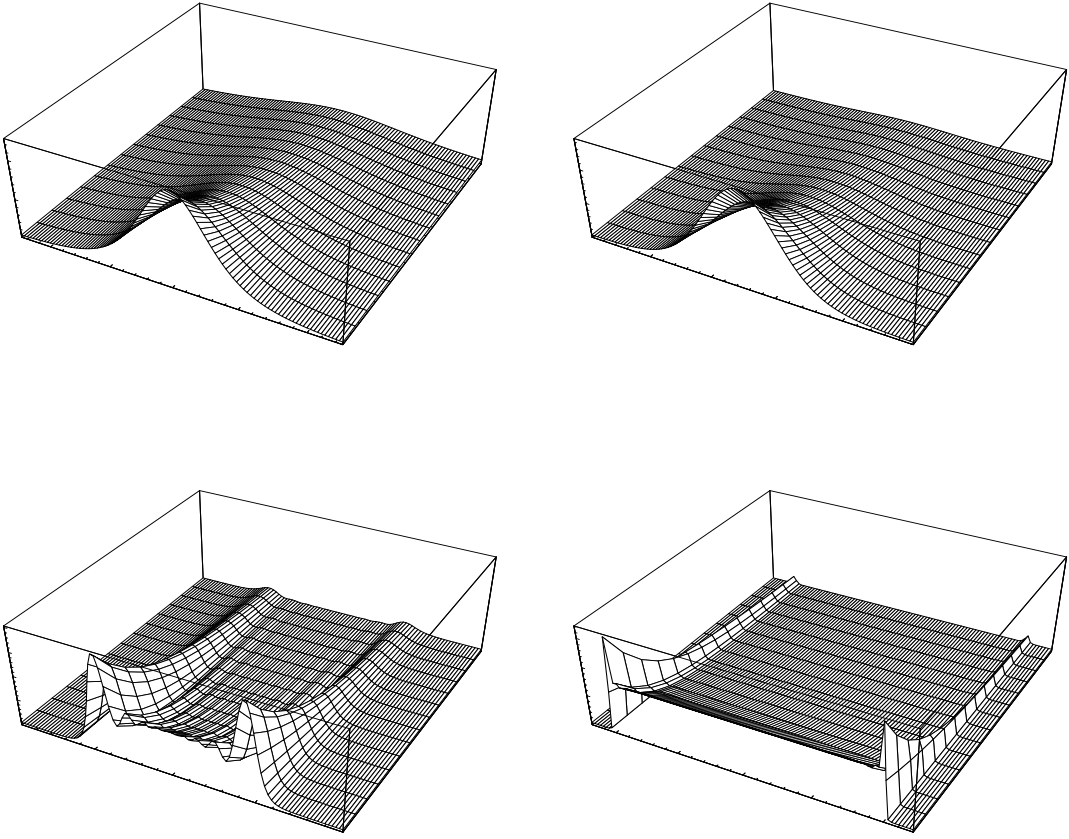}}
\caption{{\it The momentum distribution, $n(-\bp )/2$, of the positrons at zero
temperature. All quantities are expressed in units of the electric
field.\label{figf1}}}
\end{figure}
Writing,
\begin{equation}
\Om ^0=\chi _1^0 (\chi _1^0)^\da +\chi _2^0 (\chi _2^0)^\da -1, \; \;
\Om ^g=\chi _1^g (\chi _1^g)^\da +\chi _2^g (\chi _2^g)^\da
-1,\label{eq:spinors}
\end{equation}
\begin{figure}[t]
\centerline{\psfig{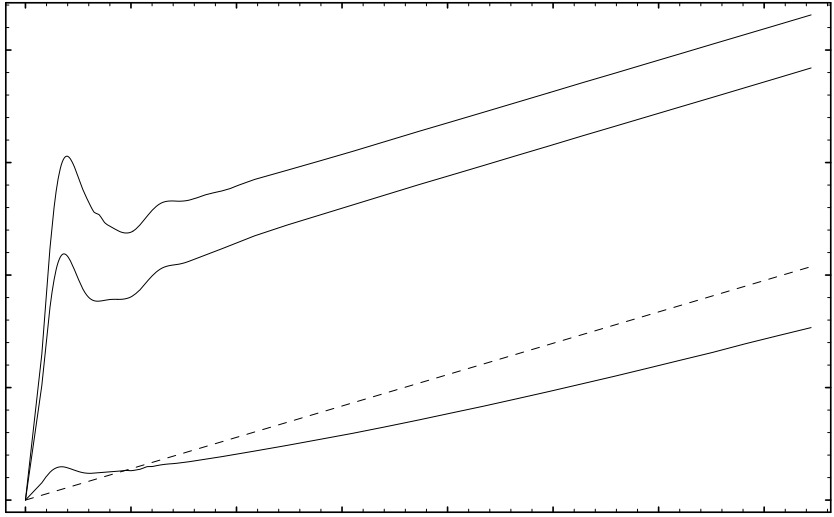}}
\vspace{1cm}
\caption{{\it The average number of created fermions for $\barm =1$. The dashed
line shows
the asymptotic behaviour at zero temperature. All quantities are expressed in
units of the electric field.}}
\label{figf2}
\end{figure}
one finds,
\begin{equation}
\mbox{tr}(\Om ^0 \Om ^g)=\sum _{i,j=1}^2| (\chi ^0_i ,\chi ^g _j)|^2-4,
\end{equation}
where $(\chi ,\lambda )=\chi _\alpha \lambda ^* _\alpha $ and having used that
the
norm $(\chi ,\chi )$ is conserved in time. Using \refpa{eq:chi1t} and
\refpa{eq:chi2t}
we obtain,
\begin{equation}
\sum _{i,j=1}^2| (\chi ^0_i ,\chi ^g _j)|^2=2 | (\chi ^0_1 ,\chi ^g _1)|^2
=\frac{2}{\barom _0 (\barom _0+\barp ^3)
\barom _0^g (\barom _0^g +\barp ^3_g)} |\Lambda y_1^g+(\barom _0+\barp ^3)
(\barom _0^g +\barp ^3_g) y_2^g |^2.
\end{equation}
Finally, writing $y_1=r e^{i \theta}$ and again using the conservation of the
spinor
norm, we find
\begin{equation}
\av {\hat N} ^{\rm cr} (t _f) =V \int _p n(\bp ,t_f)
\tanh (\frac{\bar{\beta} \barom _0^g}{2}) ,\label{eq:nint}
\end{equation}
having defined the zero temperature number density $n(\bp ,t_f)$ as
\begin{equation}
n(\bp ,t_f) =2 \left\lbrack 1-\frac{\barp ^3}{\barom _0}-(1-\frac{\barp
^3_g}{\barom _0^g})
\frac{l_g}{\barom _0} \right\rbrack ,\label{eq:nberdens}
\end{equation}
where
\begin{eqnarray}
\ddot r-\frac{l^2}{r^3}+\barom ^2 r &=& 0,\; \; r(0)=1,\; \; \dot r(0)=0,\\
\dot l &=& r^2,\; \; l(0)=\barom _0,\label{eq:l}\\
\dot \theta  &=& \frac{l}{r^2},\; \; \theta (0)=0 .
\end{eqnarray}
The subscript (superscript) $g$ still indicates gauge transformed quantities
i.e.
$\barp ^3$ is shifted, $\barp ^3_g=\barp ^3+\bart _f$.
Figure~\ref{figf1} shows $n(-\bp,t _f)/2$, the average momentum distribution of
the
created positrons at zero temperature. Comparing with the corresponding result
for
bosons, figure~\ref{figb1}, there is an interesting difference for small times.
Lots
of fermions are created with zero momentum while hardly any bosons are. The
result of
numerically evaluating the integral
\refpa{eq:nint} with a mass $\barm =1$ for some different temperatures is shown
in
figure~\ref{figf2}. Increasing the temperature clearly decreases the particle
production but again, as for bosons, for large enough times the slope is
independent
of temperature. This suppression of particle production can be understood
in a simple model at fixed spin and momentum. We then have four states:
$\ket{0},
\ket{e^-}, \ket{e^+}$ and $\ket{e^{-}e^{+}}$. Due to charge conservation the
only
allowed particle production- and annihilation processes are:
$\ket{0}\ra\ket{e^{-}e^{+}}~,~\ket{e^{-}e^{+}}\ra\ket{0}$.
At infinite temperature the probabilities for $\ket{0}$ and $\ket{e^-e^+}$ are
equal in
the initial ensemble.
Hence creation and annihilation processes are equally probable leading to no
creation
on the average.

If we have an electric field with an arbitrary time dependence as for bosons,
\refpa{e:at}, we have to change $\barom$ accordingly,
gauge transformations become $\barp ^3_g=\barp ^3+F(\bart _f)$ and we also have
to change
\refpa{eq:l} into $\dot l=f(\bart ) r^2$.

\subsection{Asymptotic expansion}
Let us now examine what happens to the number density $n(\Lambda ,\barp
^3,\bart _f)$
when $\bart _f$ is large, $\bart _f >>1$.
As for bosons, consider $\Lambda $ and
$\barp ^3$ satisfying,
\begin{equation}
\sqrt{\Lambda} << \bart _f, \; \; \; \; \; \sqrt{\Lambda} -\bart _f << \barp ^3
<< -\sqrt{\Lambda} ,
\; \; \; \; \; \barp ^3 << -1 \label{eq:int}.
\end{equation}
The initial condition for $\Om ^g$ is,
\begin{equation}
\Om ^g (\bp ,0)=-\gamma ^0 \frac{( \bfga \cdot (\bp +e {\bf A}(t_f))+m)}{\om _0
(\bp
+e {\bf A}(t_f))} .
\end{equation}
Using \refpa{e:asymp1} one finds $\Om ^g (\bp ,t)$ for $\bp $ restricted by
\refpa{eq:int}
as,
\begin{eqnarray}
\chi _1^g (\bart )&=& \sqrt{2} e^{\frac{-\pi \Lambda }{8}} \left\lbrack
\begin{array}{l}
\frac{1+i}{2} \barm D_{-\frac{i}{2} \Lambda -1} (-(1+i) (\bart-\bart _f-\barp
^3)) \\
0 \\
D_{-\frac{i}{2} \Lambda} (-(1+i) (\bart-\bart _f-\barp ^3)) \\
\frac{1+i}{2}(\barp ^1+i \barp ^2) D_{-\frac{i}{2} \Lambda -1} (-(1+i)
(\bart-\bart _f-\barp ^3))  \end{array} \right\rbrack, \\
\chi _2^g (\bart )&=& \sqrt{2} e^{\frac{-\pi \Lambda }{8}} \left\lbrack
\begin{array}{l}
\frac{1+i}{2}(-\barp ^1+i \barp ^2) D_{-\frac{i}{2} \Lambda -1} (-(1+i)
(\bart-\bart _f-\barp ^3)) \\
D_{-\frac{i}{2} \Lambda} (-(1+i) (\bart-\bart _f-\barp ^3)) \\
0 \\
\frac{1+i}{2} \barm D_{-\frac{i}{2} \Lambda -1} (-(1+i) (\bart-\bart _f-\barp
^3))
\end{array} \right\rbrack  .
\end{eqnarray}
Note that the $\chi _j^g$ here are not normalized exactly in the same way as in
\refpa{eq:c10} and \refpa{eq:c20}, the normalization differs by some irrelevant
phases.
Again for $\bp$ restricted by \refpa{eq:int}, we obtain, using
\refpa{e:asymp2},
\begin{equation}
n(\Lambda ,\barp ^3,\bart _f)=2-\frac{1}{2}\mbox{tr}(\Om ^0 \Om ^g (t_f))=
4 e^{-\pi \Lambda } .
\end{equation}
Thus, using a similar reasoning as for bosons, we conclude that for $\bart _f
>>1$
and $\bart _f >> \barm$,
\begin{equation}
n(\Lambda ,\barp ^3,\bart _f)=\left\{ \begin{array}{ll}
4 e^{-\pi \Lambda}, \; \; \; & -\bart _f < \barp ^3 < 0, \\
0, \; \; \;  & \mbox{otherwise} .\label{eq:ntf}
\end{array} \right.
\end{equation}
Comparing to the analogous expression for bosons, \refpa{e:numberdensity3},
we note that
the only difference is a factor of $2$ arising from the spin degrees of freedom
of
the fermions. Accordingly we obtain
\begin{equation}
\av {\hat N} ^{\rm cr} (\bart _f) =\frac{{\bar V} }{2 (2\pi )^2} \int _{-\bart
_f} ^0
d\barp ^3 \int _{\barm ^2} ^\infty d\Lambda \; 4\tanh (\frac{{\bar \beta}
\barom _0^g }{2})
e^{-\pi \Lambda }, \; \; \; \; \bart _f >>1, \; \; \; \bart _f >> \barm
.\label{eq:asy}
\end{equation}
Considering the limits $\frac{1}{\barbe } << \bart _f $ and $\frac{1}{\barbe}
>>
\bart _f$ of \refpa{eq:asy}, we find
\begin{eqnarray}
\av {\hat N} ^{\rm cr} (\bart _f) &=& \frac{4{\bar V}}{(2\pi )^3} \bart _f
e^{-\pi \barm ^2}=\frac{2 \alpha }{\pi^2 }VE^2 t_f  e^{-\frac{\pi m^2}{|e E|}}
, \; \; \; \; \frac{1}{\barbe } << \bart _f , \\
\av {\hat N} ^{\rm cr} (\bart _f)  &=& \frac{\barbe {\bar V} }{(2\pi )^3}
\bart _f ^2 e^{-\pi \barm ^2}=(\frac{\alpha }{\pi})^{3/2} \beta V |E|^3 t_f ^2
e^{-\frac{\pi m^2}{|e E|}}, \; \; \; \; \frac{1}{\barbe} >> \bart _f .
\end{eqnarray}

\subsection{Probabilities}
We will now as for bosons give the expressions for the probability $P(n,m;t
_f)$ of
finding $n$ electrons and $m$ positrons in the ensemble at time $t=t_f$. The
calculations are however a good deal more involved for fermions so we will
illustrate
what happens by calculating $P(0,0;t_f)$ explicitly. A particular choice of
groundstate $\Psi _0$
is, (remember that the representation is reducible),
\begin{equation}
\Psi _0 (\eta ^* , \eta )=\det (\frac{I}{\sqrt{2}}) \exp (\int_p\eta ^* \Om ^0
\eta) .
\end{equation}
We find in momentum space
\begin{eqnarray}
\lefteqn{\bra {\Psi _0}\hat \rho (t_f) \ket{\Psi _0}=}\nn \\
&=& \det (\frac{I}{4})
\det (\Om ^0 +\Om ^\beta ) \times \nn \\
&& \det \left((\Om ^\beta -1)(\Om ^0+ \Om ^\beta)
(\Om ^\beta -1)^{-1}  -(\Om ^\beta -1) (\Om ^\beta +1) (\Om ^0 +\Om ^\beta
)^{-1}
 \right)
\nn \\
&=& \det \left[{\mbox {det}} _{sp} \left( \frac{
(\Om ^\beta -1) (\Om ^0 +\Om ^\beta )(\Om ^\beta +1)
(\Om ^0+\Om ^\beta )}{4(th^2-1)}+\frac{1-th^2}{4} \right) \right]
\end{eqnarray}
where $\Om ^\beta =th \; \Om ^g (t_f)$, $th=\tanh (\frac{ \barbe \barom _0
^g}{2})$ and
${\mbox {det}} _{sp}$ denotes the determinant evaluated only with respect to
the
spinor indices. Now express $\Om ^0$ and $\Om ^g (t_f)$ in terms of spinors,
see \refpa{eq:spinors}, and use
\begin{equation}
{\mbox {det}} _{sp}(A)=\frac{1}{4!}\lbrack (\mbox{tr} A)^4-6(\mbox{tr} A)^2
(\mbox{tr} A^2)+8 \mbox{tr} A \; \mbox{tr} A^3+3 (\mbox{tr} A^2)^2-6 \mbox{tr}
A^4 \rbrack,
\end{equation}
which holds for a $4 \times 4$-matrix $A$. Using a computer to handle the
algebra we
obtain
\begin{eqnarray}
P(0,0;t_f ) &=& \bra {\Psi _0}\hat \rho (t_f) \ket{\Psi _0} = \det
(\frac{1}{16} (1+th \; (| (\chi _1 ^0,
\chi _1^g)|^2-2)+th^2)^2)\nn \\
&=& \exp (2V \int _p \ln \lbrack \frac{1}{4} (1+th (2-n(t_f))+th^2) \rbrack ),
\hspace{5cm} \label{eq:pzero}
\end{eqnarray}
expressed in terms of the zero temperature number density $n(t_f)$ defined in
\refpa{eq:nberdens} .
At zero temperature this reduces to
\begin{equation}
P(0,0;t_f)=\exp \left\lbrack 2 \Tr \ln (1-\frac{n(t_f)}{4}) \right\rbrack ,
\end{equation}
which for large times, $\bart _f >> 1$ , by \refpa{eq:ntf} becomes
\begin{equation}
P(0,0; t_f)=\exp \left\lbrack \frac{\barV \bart _f}{(2 \pi )^2}
\int _{\barm ^2} ^\infty  d\Lambda \ln (1-e^{-\pi \Lambda }) \right\rbrack.
\end{equation}
Again this result agrees with the result of Schwinger \cite{schwinger}. In
general we
express the probability $P(n,m;t_f)$ in terms of a generating function
$f(x,y;t_f)$
as for bosons, \refpa{e:prob},
\begin{equation}
f(x,y;t_f)=\exp( 2 \Tr \ln \lbrack (1+x \; g(t_f))(1+y \; g(t_f))+x\;
y \; h(t_f) \rbrack ),
\end{equation}
where $g(t_f)$ and $h(t_f)$ are
\begin{eqnarray}
g(t_f) &=& \frac{2}{(2-n(t_f)+2 \coth \barbe \barom _0 ^g) \sinh \barbe \barom
_0 ^g},\nn \\
h(t_f) &=& \frac{n(t_f) (4-n(t_f))}{(2-n(t_f)+2 \coth \barbe \barom _0 ^g)^2} .
\end{eqnarray}
Finally calculating the expectation value of the number operator
\begin{eqnarray}
\av{\hat N}(t_f) &=& \sum _{n,m}^\infty (n+m) P(n,m)=f(1,1)^{-1} \pa _x f(x,x)
| _{x=1}\nn \\
&=& \Tr (2+\tanh (\frac{ \barbe \barom _0 ^g}{2}) (n(t_f)-2)) ,
\end{eqnarray}
which agrees with \refpa{eq:exnop}.

\section{Final remarks}

We will briefly comment on the infinite time limit of $P(0,0;\tf)$ since this
quantity
is usually discussed. We will do this for fermions. For bosons,
a similar argument may be made.
The expression \refpa{eq:pzero} for $P(0,0,\tf)$ may be rewritten when
$\bartf>>1$.
Using
\refpa{eq:ntf} we find for an arbitrary temperature
\be
P(0,0;\tf)=\exp\left[C(\barbeta) +
\frac{2\barV}{2(2\pi)^2}\int_{\barm^2}^{\infty} d\Lambda \int _{-\bartf}^{0}
d\barp^3 \ln \frac{1+th
(2-4e^{-\pi \Lambda}) +th^2}{(1+th)^2}\right]
\ee
where $C(\barbeta)$ is a time independent constant and
$th=\tanh (\frac{\barbeta \barom _0^g}{2})$. The only time dependence is in
 $\barom_0^g(\tf)$. From this expression it is clear that $w$, in the
literature misleadingly known as the pair production rate, is
\be
w=\lim_{t_f \ra\infty}\frac{\ln P(0,0;\tf)}{V\tf}=
\frac{\alpha E^2}{\pi}\int_{\barm^2}^{\infty} d\Lambda
 \ln(1-e^{-\pi \Lambda})
\ee
for any temperature. Therefore in the strict infinite time limit $w$ is
independent of
temperature, even though for intermediate times it certainly does depend on
temperature. However, the quantity that is directly related to the observable
particle
production is not $w$ but the average of the number operator, $\av{\hat N}^{\rm
cr}$,
which we have thoroughly discussed.
\vspace{1cm}

\centerline{\bf{Acknowledgements}}
We wish to thank Per Sundell for some help in the early stages of this work,
Per Salomonsson for a stimulating discussion and especially Bo-Sture Skagerstam
for
many interesting remarks and useful comments throughout this work.

\end{document}